  \documentclass{gretsi}

  \usepackage{times}
 \usepackage[english,francais]{babel} 
 \usepackage[T1]{fontenc}
 \usepackage{ae,aecompl} 

\usepackage{graphicx} 
\graphicspath{{Eps/}}

\titre{Approche variationnelle pour le calcul bay\'esien dans les probl\`emes inverses en imagerie}
\auteur{\coord{Ali}{Mohammad-Djafari}{1}}
\adresse{\affil{1}{Laboratoire des signaux et syst\`emes \\
         (UMR 08506 du CNRS, Sup\'elec et Univ Paris Sud), \\ 
        Sup\'elec, Plateau de Moulon, 91192 Gif-sur-Yvette Cedex, France}}
\email{djafari@lss.supelec.fr}

\resumefrancais{Dans une approche bay\'esienne non supervis\'ee pour la r\'esolution d'un probl\`eme 
inverse, on cherche \`a estimer conjointement la grandeur inconnue $\fb$ et 
les param\`etres $\thetab$. Ceci se fait en utilisant la loi 
\apost conjointe $p(\fb,\thetab|\gb)$. L'expression de cette loi est souvent complexe 
et son exploration et le calcul des estimateurs bay\'esiens n\'ecessitent soit l'optimisation 
des crit\`eres souvent non convexes ou le calcul d'esp\'erances des lois non gaussiennes 
multivari\'ees. Dans tous ces cas, il y a souvent besoin de faire des approximations. 
Nous avions d\'ej\`a explor\'e les possibilit\'es de l'approximation de Laplace et les 
m\'ethodes d'\'echantillonnage MCMC. Ici, nous explorons l'approximation de 
$p(\fb,\thetab|\gb)$ par une loi s\'eparable en $\fb$ et en $\thetab$. Ceci permet 
de proposer des algorithmes it\'eratifs plus abordables en co\^ut de calcul, surtout, 
si on choisit ces lois approchantes dans des familles des lois exponentielles. 
Le principal objet de ce papier est de fournir des d\'etails des diff\'erents algorithmes 
que l'on obtient pour diff\'erents choix de ces familles.}

\resumeanglais{In a non supervised Bayesian estimation approach for inverse problems in imaging systems, 
one tries to estimate jointly the unknown image pixels $\fb$ and the hyperparameters $\thetab$. 
This is, in general, done through the joint posterior law $p(\fb,\thetab|\gb)$. The expression of this joint law is often very complex and its exploration through sampling and computation of the point estimators such as 
MAP and posterior means need either optimization of non convex criteria or int\'egration of 
non Gaussian and multi variate probability laws. In any of these cases, we need to do approximations. 
We had explored before the possibilities of Laplace approximation and sampling by MCMC. 
In this paper, we explore the possibility of approximating this joint law by a separable one in $\fb$ 
and in $\thetab$. This gives the possibility of developing iterative algorithms with more reasonable 
computational cost, in particular, if the approximating laws are choosed in the exponential conjugate 
families. The main objective of this paper is to give details of different algorithms we obtain with 
different choices of these families.}

\begin{document}
\def\bdoc{\begin{document}}             \def\edoc{\end{document}}
\def\babs{\begin{abstract}}             \def\eabs{\end{abstract}}
\def\bcc{\begin{center}}                \def\ecc{\end{center}}
\def\ben{\begin{enumerate}}             \def\een{\end{enumerate}}
\def\bit{\begin{itemize}}               \def\eit{\end{itemize}}
\def\bpic{\begin{picture}}              \def\epic{\end{picture}}
\def\beq{\begin{equation}}              \def\eeq{\end{equation}}
\def\barr{\begin{array}}                \def\earr{\end{array}}
\def\btab{\begin{tabular}}              \def\etab{\end{tabular}}
\def\btabu{\begin{tabular}}             \def\etabu{\end{tabular}}
\def\bcc{\begin{center}}                \def\ecc{\end{center}}
\def\beqn{\begin{eqnarray}}             \def\eeqn{\end{eqnarray}}
\def\beqnx{\begin{eqnarray*}}           \def\eeqnx{\end{eqnarray*}}
\def\bfig{\begin{figure}}               \def\efig{\end{figure}}
\def\bver{\begin{verb}}						 \def\ever{\end{verb}}

\def\d#1{\,\mbox{d} #1}
\def\disp#1{\displaystyle{#1}}

\def\iii{\int_{-\infty}^{+\infty}}
\def\izi{\int_{0}^{\infty}}
\def\izpi{\int_{0}^{\pi}}
\def\izdpi{\int_{0}^{2\pi}}
\def\intd{\int\kern-.8em\int}
\def\intt{\int\kern-.8em\int\kern-.8em\int}
\def\intg{\int\kern-1.1em\int}

\def\xh{\widehat{x}}
\def\thetah{\widehat{\theta}}
\def\betah{\widehat{\beta}}

\def\xbh{\widehat{\xb}}
\def\thetabh{\widehat{\thetab}}
\def\betabh{\widehat{\betab}}

\def\xbhk{\widehat{\xb}^{k}}
\def\thetahk{\widehat{\theta}^{k}}
\def\betahk{\widehat{\beta}^{k}}

\def\xbhkp{\widehat{\xb}^{k+1}}
\def\thetahkp{\widehat{\theta}^{k+1}}
\def\betahkp{\widehat{\beta}^{k+1}}

\def\thetabhk{\widehat{\thetab}^{k}}
\def\betabhk{\widehat{\betab}^{k}}

\def\thetabhkp{\widehat{\thetab}^{k+1}}
\def\betabhkp{\widehat{\betab}^{k+1}}

\def\thetamin{\theta_{\mbox{\tiny min}}}
\def\thetamax{\theta_{\mbox{\tiny max}}}
\def\betamin{\beta_{\mbox{\tiny min}}}
\def\betamax{\beta_{\mbox{\tiny max}}}

\def\ra{\rightarrow}
\def\la{\leftarrow}
\def\da{\downarrow}
\def\ua{\uparrow}

\def\Ra{\Rightarrow}
\def\La{\Leftarrow}
\def\Da{\Downarrow}
\def\Ua{\Uparrow}

\def\lra{\longrightarrow}
\def\lla{\longleftarrow}
\def\Lra{\Longrightarrow}
\def\Lla{\longleftarrow}

\def\lrarr{\leftrightarrow}
\def\Lrarr{\Leftrightarrow}
\def\udarr{\updownarrow}
\def\Uparr{\Updownarrow}

\def\expf#1{\exp\left[ {#1} \right]}

\def\argmax#1#2{\arg\max_{#1}\left\{ #2 \right\}}
\def\argmin#1#2{\arg\min_{#1}\left\{ #2 \right\}}

\def\Prob#1{\mbox{Pr}\left\{#1\right\}}
\def\var#1{\mbox{Var}\left\{#1\right\}}
\def\cov#1{\mbox{Cov}\left\{#1\right\}}
\def\corr#1{\mbox{Corr}\left\{#1\right\}}
\def\trace#1{\mbox{Tr}\left\{#1\right\}}
\def\rang#1{\mbox{rang}\left\{#1\right\}}
\def\det#1{\mbox{d\'et}\left\{#1\right\}}

\def\gbu{\underline{\gb}}
\def\fbu{\underline{\fb}}
\def\zbu{\underline{\zb}}
\def\epsilonbu{\underline{\epsilonb}}
\def\thetabu{\underline{\thetab}}

\def\sigmae{{\sigma_{\epsilon}}}
\def\Sigmae{{\Sigmab_{\epsilonb}}}
\def\Sigmaf{{\Sigmab_{\fb}}}
\def\Sigmag{{\Sigmab_{\gb}}}

\def\fbuh{\widehat{\fbu}}
\def\zbuh{\widehat{\fbu}}
\def\thetabuh{\widehat{\thetabu}}

\def\fbh{\widehat{\fb}}
\def\Pbh{\widehat{\Pb}}
\def\Sigmabh{\widehat{\Sigmab}}
\def\Abh{\widehat{\Ab}}
\def\Bbh{\widehat{\Bb}}
\def\thetabh{\widehat{\thetab}}
\def\Rbe{\Rb_{\epsilon}}
\def\Rbeh{\widehat{\Rbe}}

\def\Rjk{{R_j}_k}
\def\fbik{{\fb_i}_k}
\def\fbjk{{\fb_j}_k}

\def\mik{{m_i}_k}
\def\sigmaik{{\sigma_i^2}_k}
\def\Sigmaik{{\Sigmab_i}_k}
\def\alphaik{{\alpha_i}_k}
\def\betaik{{\beta_i}_k}

\def\muik{{\mu_i}_k}
\def\vik{{v_i}_k}

\def\mjk{{m_j}_k}
\def\sigmajk{{\sigma_j^2}_k}
\def\Sigmajk{{\Sigmab_j}_k}
\def\alphajk{{\alpha_j}_k}
\def\betajk{{\beta_j}_k}

\def\mujk{{\mu_j}_k}
\def\vjk{{v_j}_k}
\def\njk{{n_j}_k}

\def\alphae{\alpha^{\epsilon}}
\def\betae{\beta^{\epsilon}}
\def\alphaez{\alpha^{\epsilon}_{0}}
\def\betaez{\beta^{\epsilon}_{0}}
\def\alphaei{\alpha^{\epsilon}_{i}}
\def\betaei{\beta^{\epsilon}_{i}}
\def\alphaeiz{\alpha^{\epsilon}_{i0}}
\def\betaeiz{\beta^{\epsilon}_{i0}}

\def\mbz{{{\mb}_z}}
\def\Sigmabz{{{\Sigmab}_z}}
\def\diag#1{\mbox{diag}\left[#1\right]}

\def\sigmah{\widehat{\sigma}}
\def\fh{\widehat{f}}
\def\sigmaz{\sigma_z}

\def\blue#1{{\color{blue}#1}}
\def\red#1{{\color{red}#1}}
\def\green#1{{\color{green}#1}}

\def\sumi{\sum_{i=1}^{M} }
\def\sumj{\sum_{j=1}^{N} }
\def\lambdah{\wh{\lambda}}
\def\lambdabh{\wh{\lambdab}}

\def\det#1{\hbox{det}\left(#1\right)}
\def\defined{\,\shortstack{$\triangle$\\ =}\,}

\def\zmat{\left[\matrix{0}\right]}
\def\det#1{\hbox{det}(#1)}
\def\th{\widehat{t}}
\def\xh{\widehat{x}}
\def\lambdah{\widehat{\lambda}}
\def\muh{\widehat{\mu}}
\def\sigmah{\widehat{\sigma}}
\def\rhoh{\widehat{\rho}}
\def\betah{\widehat{\beta}}
\def\alphah{\widehat{\alpha}}
\def\tbh{\widehat{\tb}}
\def\mubh{\widehat{\mub}}
\def\sigmabh{\widehat{\sigmab}}
\def\rhobh{\widehat{\rhob}}
\def\oneb{\mbox{\bf 1}}

\def\bm#1{\mbox{\boldmath $#1$}}
\def\zerob{{\bf 0}}
\def\oneb{{\bf 1}}
\def\intg{\int\kern-1.1em\int}
\def\expf#1{\exp\left[ {#1} \right]}

\def\gbu{\underline{\gb}}
\def\fbu{\underline{\fb}}
\def\zbu{\underline{\zb}}
\def\epsilonbu{\underline{\epsilonb}}
\def\uthetab{\underline{\thetab}}

\def\sigmae{{\sigma_{\epsilon}}}
\def\Sigmae{{\Sigma_{\epsilonb}}}
\def\Sigmabe{{\Sigmab_{\epsilonb}}}
\def\Sigmaf{{\Sigmab_{\fb}}}
\def\Sigmag{{\Sigmab_{\gb}}}

\def\fbuh{\widehat{\fbu}}
\def\zbuh{\widehat{\fbu}}
\def\uthetabh{\widehat{\uthetab}}

\def\fbh{\widehat{\fb}}
\def\Sigmabh{\widehat{\Sigmab}}
\def\Abh{\widehat{\Ab}}
\def\thetabh{\widehat{\thetab}}

\def\Rjk{{R_j}_k}
\def\fbik{{\fb_i}_k}
\def\fbjk{{\fb_j}_k}

\def\mik{{m_i}_k}
\def\sigmaik{{\sigma_i^2}_k}
\def\Sigmaik{{\Sigmab_i}_k}
\def\alphaik{{\alpha_i}_k}
\def\betaik{{\beta_i}_k}

\def\muik{{\mu_i}_k}
\def\vik{{v_i}_k}

\def\mjk{{m_j}_k}
\def\sigmajk{{\sigma_j^2}_k}
\def\Sigmajk{{\Sigmab_j}_k}
\def\alphajk{{\alpha_j}_k}
\def\betajk{{\beta_j}_k}

\def\mujk{{\mu_j}_k}
\def\vjk{{v_j}_k}
\def\njk{{n_j}_k}

\def\alphae{\alpha^{\epsilon}}
\def\betae{\beta^{\epsilon}}
\def\alphaez{\alpha^{\epsilon}_{0}}
\def\betaez{\beta^{\epsilon}_{0}}
\def\alphaei{\alpha^{\epsilon}_{i}}
\def\betaei{\beta^{\epsilon}_{i}}
\def\alphaeiz{\alpha^{\epsilon}_{i0}}
\def\betaeiz{\beta^{\epsilon}_{i0}}
\def\alphaeiz{\alpha^{\epsilon}_{i0}}
\def\betaeiz{\beta^{\epsilon}_{i0}}
\def\alphaeiz{\alpha^{\epsilon}_{i0}}
\def\betaeiz{\beta^{\epsilon}_{i0}}

\def\mbz{{{\mb}_z}}
\def\Sigmabz{{{\Sigmab}_z}}

\def\vect{\mbox{vect}}
\def\lra{\longrightarrow}

\def\gir{g_i(\rb)}
\def\fjr{f_j(\rb)}
\def\zjr{z_j(\rb)}
\def\epsilonir{\epsilon_i(\rb)}
\def\gbr{\gb(\rb)}
\def\fbr{\fb(\rb)}
\def\zbr{\zb(\rb)}
\def\epsilonbr{\epsilonb(\rb)}
\def\fbhr{\fbh(\rb)}
\def\Sigmabhr{\Sigmabh(\rb)}

\def\mbzr{\mb_{z(\rb)}}
\def\Sigmabzr{\Sigmab_{z(\rb)}}
\def\Sigmagbzr{{\Sigmab_g}_{z(\rb)}}

\def\mjzjr#1{{m_{#1}}_{z_{#1}(\rb)}}
\def\sigmajzjr#1{{\sigma_{#1}}_{z_{#1}(\rb)}}

\def\Beta{B}
\def\Betab{\bm{\Beta}}
\def\Betabe{\bm{\Beta}_{\epsilonb}}

\def\fh{\widehat{f}}
\def\sigmah{\widehat{\sigma}}
\def\sigmaei{\sigma_{\epsilon_i}^2}

\def\thetah{\widehat{\theta}}
\def\sigmah{\widehat{\sigma}}

\def\Ftxm{\widetilde{F}_{X}(x_-)}
\def\Ftxp{\widetilde{F}_{X}(x_+)}

\def\NU{{\cal V}}

\def\tFx{\widetilde{F}_{X}}

\def\Fx{F_{X}(x)}
\def\fx{f_{X}(x)}

\def\Fxv{F_{X}(x)}
\def\fxv{f_{X}(x)}

\def\Fxi{{F}_{X_i}(x_i)}
\def\fxi{{f}_{X_i}(x_i)}

\def\Gnu{G_{\NU}(\nu)}
\def\gnu{g_{\NU}(\nu)}

\def\Fnu{F_{\NU}(\nu)}
\def\fnu{f_{\NU}(\nu)}

\def\Finnu#1{F^{-1}_{\NU}(#1)}

\def\Gnua#1{G_{\NU}(#1)}
\def\Gnub#1{G_{\NU}^{-1}(#1)}
\def\Gnuh{G_{\NU}^{-1}(1/2)}

\def\Ftx{\widetilde{F}_{X}(x)}
\def\ftx{\widetilde{f}_{X}(x)}

\def\Ftxi{\widetilde{F}_{X_i}(x_i)}
\def\ftxi{\widetilde{f}_{X_i}(x_i)}

\def\Ftxv{\widetilde{F}_{X}(x)}
\def\FtX#1{\widetilde{F}_{X}(#1)}
\def\ftxv{\widetilde{f}_{X}(x)}

\def\fxnu{f_{X|\NU}(x|\nu)}
\def\fxNU{f_{X|\NU}(x|\NU)}
\def\FxNU{F_{X|\NU}(x|\NU)}

\def\fxNUtheta{f_{X|\NU,\theta}(x|\NU,\theta)}
\def\FxNUtheta{F_{X|\NU,\theta}(x|\NU,\theta)}

\def\Fxnu{F_{X|\NU}(x|\nu)}
\def\Fxnui#1{F_{X|\NU}(x|\nu_#1)}

\def\FxN#1{F_{X|\NU}(#1)}
\def\FxNU{F_{X|\NU}(x|\NU)}
\def\Fxnun#1{F_{X|\NU}(x|#1)}
\def\Fxinun#1{F_{X_i|\NU}(x_i|#1)}
\def\fxnu{f_{X|\NU}(x|\nu)}
\def\fxNU{f_{X|\NU}(x|\NU)}

\def\Fxvnu{F_{X|\NU}(x|\nu)}
\def\Fxvnun#1{F_{X|\NU}(x|#1)}
\def\FxvNU{F_{X|\NU}(x|\NU)}

\def\FXvnu{F_{X|\NU}}

\def\fxvnu{f_{X|\NU}(x|\nu)}
\def\fxvNU{f_{X|\NU}(x|\NU)}

\def\FXvNU#1{{F}_{X|\NU}(#1|\NU)}
\def\FXvn#1{{F}_{X|\NU}(x|#1)}

\def\Fxtheta{F_{X|\theta}(x|\theta)}
\def\fxtheta{f_{X|\theta}(x|\theta)}

\def\Fxvtheta{F_{X|\theta}(x|\theta)}
\def\fxvtheta{f_{X|\theta}(x|\theta)}

\def\Ftxtheta{\widetilde{F}_{X|\theta}(x|\theta)}
\def\ftxtheta{\widetilde{f}_{X|\theta}(x|\theta)}

\def\Ftxvtheta{\widetilde{F}_{X|\theta}(x|\theta)}
\def\ftxvtheta{\widetilde{f}_{X|\theta}(x|\theta)}

\def\Fxnutheta{F_{X|\NU,\theta}(x|\nu,\theta)}
\def\fxnutheta{f_{X|\NU,\theta}(x|\nu,\theta)}

\def\ftheta{f_{\Theta}(\theta)}
\def\fthetaxnuz{f_{\Theta|X,\nu_0}(\theta|x,\nu_0)}
\def\fthetax{f_{\Theta|X}(\theta|x)}

\def\Fxvnutheta{F_{X|\NU,\theta}(x|\nu,\theta)}
\def\FxvNUtheta{F_{X|\NU,\theta}(x|\NU,\theta)}
\def\fxvnutheta{f_{X|\NU,\theta}(x|\nu,\theta)}
\def\Ftxvnutheta{\widetilde{F}_{X|\theta}(x|\theta)}
\def\ftxvnutheta{\widetilde{f}_{X|\theta}(x|\theta)}

\def\FxvNUm{F_{X|\NU}(x_-|\NU)}
\def\FxvNUp{F_{X|\NU}(x_+|\NU)}

\def\Fxinu{F_{X_i|\NU}(x_i|\nu)}
\def\fxinu{f_{X_i|\NU}(x_i|\nu)}

\def\Ftxi{\widetilde{F}_{X_i}(x_i)}
\def\ftxi{\widetilde{f}_{X_i}(x_i)}

\def\fxitheta{f_{X_i|\theta}(x_i|\theta)}
\def\ftxitheta{\widetilde{f}_{X_i|\theta}(x_i|\theta)}

\def\thetah{\widehat{\theta}}
\def\thetat{\widetilde{\theta}}

\def\lnuxv{l_{\nu|X}(\nu|x)}

\def\Lnuxva#1{L_{\nu|X}(#1 | x)}
\def\lnuxva#1{l_{\nu|X}(#1 | x)}

\def\Prob#1{\mbox{P}\left(#1\right)}

\def\lthetaxvnuz{l(\theta)}

\def\lthetaxvnuz{l_{\theta|x,\nu_0}(\theta | x,\nu_0)}
\def\fxnuztheta{f_{X|\nu_0,\theta}(x | \nu_0,\theta)}
\def\fxinuztheta{f_{X|\nu_0,\theta}(x_i | \nu_0,\theta)}
\def\fxvtheta{f_{X|\theta}(x | \theta)}
\def\lthetaxv{l_{\theta|X}(\theta | x)}
\def\ltthetaxv{\widetilde{l}_{\theta|X}(\theta | x)}

\def\ltheta{l(\theta )}
\def\lttheta{\widetilde{l}(\theta)}
\def\Lnu{L(\nu )}
\def\Lnui#1{L(\nu_#1)}
\def\Linu{L_i(\nu )}
\def\Linu{L_i(\nu )}
\def\LNU{L(\NU )}
\def\Lin#1{L^{-1}(#1)}
\def\L#1{L(#1)}
\def\Li#1{L_i(#1)}
\def\FNU#1{F_\NU(#1)}
\def\FinNU#1{F^{-1}_\NU(#1)}

\def\ftthetax{\tilde{f}_{\Theta|X}(\theta|x)}
\def\ER{R}

\def\Ndd#1#2#3{{\cal N}\left( #1 ;~ #2, #3 \right)}
\def\Cdd#1#2#3{{\cal C}\left( #1 ;~ #2, #3 \right)}
\def\Sdd#1#2#3#4{{\cal S}\left( #1 ;~ #2, #3, #4 \right)}
\def\Edd#1#2{{\cal E}\left( #1 ;~ #2 \right)}
\def\DEdd#1#2{{\cal DE}\left( #1 ;~ #2 \right)}
\def\Gdd#1#2#3{{\cal G}\left( #1 ;~ #2, #3 \right)}
\def\IGdd#1#2#3{{\cal IG}\left( #1 ;~ #2, #3 \right)}

\def\Xwr{X(\omega,\rb)}
\def\xwr{x(\omega,\rb)}

\def\muzw{\mu_{z}(\omega)}
\def\muzwr{\mu_{z(\rb)}(\omega)}
\def\szw{\sigma_{z}^2(\omega)}
\def\szwr{\sigma_{z(\rb)}^2(\omega)}
\def\pzw{p_{z}(\omega)}
\def\pzwr{p_{z(\rb)}(\omega)}

\def\hszw{\widehat{\sigma_{z}^2}(\omega)}
\def\hmuzw{\widehat{\mu}_{z}(\omega)}
\def\hpzw{\widehat{p}_{z}(\omega)}

\def\muzwrl{\mu_{z(\rb)}(\omega-l)}

\def\pxwrcz{p\left(\xwr ~|~ z\right)}
\def\pxwr{p\left(\xwr\right)}

\def\Xbw{\Xb(\omega)}
\def\xbw{\xb(\omega)}
\def\uXb{\underline{\Xb}}
\def\uxb{\underline{\xb}}
\def\pxbw{p(\xbw)}
\def\puxb{p(\uxb)}
\def\pzb{P(\zb)}

\def\Szw{S_z(\omega)}
\def\mubw{\mub(\omega)}

\def\Srw{S_{\rb}(\omega)}
\def\Swr{S_{\omega}(\rb)}
\def\Sbw{\Sb(\omega)}
\def\sbw{\sb(\omega)}
\def\Sbr{\Sb(\rb)}
\def\mubw{\mub(\omega)}
\def\cov#1{\mbox{cov}[#1]}

\def\pdf{\emph{pdf}}
\def\pmf{\emph{pmf}}
\def\dpdx#1#2{\frac{\partial #1}{\partial #2}}

\def\REM#1{}

\def\XXwr{X(\omega,\rb)}
\def\xxwr{x(\omega,\rb)}

\def\Xwr{X_{\omega}(\rb)}
\def\xwr{x_{\omega}(\rb)}
\def\Xrw{X_{\rb}(\omega)}
\def\xrw{x_{\rb}(\omega)}

\def\Ywr{Y_{\omega}(\rb)}
\def\ywr{y_{\omega}(\rb)}
\def\Yrw{Y_{\rb}(\omega)}
\def\yrw{y_{\rb}(\omega)}

\def\Ewr{E_{\omega}(\rb)}
\def\ewr{\epsilon_{\omega}(\rb)}
\def\Erw{E_{\rb}(\omega)}
\def\erw{\epsilon_{\rb}(\omega)}

\def\Xbr{\Xb(\rb)}
\def\xbr{\xb(\rb)}
\def\Xbw{\Xb(\omega)}
\def\xbw{\xb(\omega)}

\def\uXb{\underline{\Xb}}
\def\uxb{\underline{\xb}}
\def\uXbz{\underline{\Xb}_z}
\def\uxbz{\underline{\xb}_z}

\def\uthetab{\underline{\thetab}}
\def\uthetabz{\underline{\thetab}_z}

\def\Ywr{Y_{\omega}(\rb)}
\def\ywr{y_{\omega}(\rb)}
\def\Yrw{Y_{\rb}(\omega)}
\def\yrw{y_{\rb}(\omega)}
\def\Ybr{\Yb(\rb)}
\def\ybr{\yb(\rb)}
\def\Ybw{\Yb(\omega)}
\def\ybw{\yb(\omega)}
\def\uYb{\underline{\Yb}}
\def\uyb{\underline{\yb}}
\def\uYbz{\underline{\Yb}_z}
\def\uybz{\underline{\yb}_z}

\def\Ewr{E_{\omega}(\rb)}
\def\ewr{\epsilon_{\omega}(\rb)}
\def\Erw{E_{\rb}(\omega)}
\def\erw{\epsilon_{\rb}(\omega)}
\def\Ebr{Eb(\rb)}
\def\ebr{\epsilonb(\rb)}
\def\Ebw{Eb(\omega)}
\def\ebw{\epsilonb(\omega)}
\def\uEb{\underline{\Eb}}

\def\Erwp{E_{\rb}(\omega')}
\def\Xrwp{X_{\rb}(\omega')}

\def\xwmr{x_{\omega-1}(\rb)}
\def\muzwm{\mu_{z}(\omega-1)}

\def\rhoz{\rho_z}
\def\sez{\sigma_{\eta_z}^2}

\def\xwrp{x_{\omega}(\rb')}
\def\xrwp{x_{\rb}(\omega')}
\def\xwpr{x_{\omega'}(\rb)}
\def\xwprp{x_{\omega'}(\rb')}
\def\Xwrp{x_{\omega}(\rb')}
\def\Xwrp{X_{\omega}(\rb')}
\def\Xwpr{X_{\omega'}(\rb)}
\def\Xwprp{X_{\omega'}(\rb')}

\def\sigmaed{\sigmae^2}
\def\sigmaew{\sigmae(\omega)}
\def\sigmaedw{\sigmaed(\omega)}

\def\Sigmabe{\Sigmab_{\epsilon}}
\def\sigmabh{\widehat{\Sigmab}}
\def\xbh{\widehat{\xb}}
\def\zbh{\widehat{\zb}}
\def\qbh{\widehat{\qb}}
\def\sbh{\widehat{\sb}}

\def\argmax#1#2{\arg\max_{#1}\left\{#2\right\}}
\def\argmin#1#2{\arg\min_{#1}\left\{#2\right\}}
\def\espx#1#2{\mbox{E}_{#1}\left\{#2\right\}}
\def\esp#1{\mbox{E}\left\{#1\right\}}
\def\d#1{\mbox{~d}#1}

\def\Tr#1{\mbox{Tr}\left[#1\right]}

\def\pmatrix#1{\left[\barr{c} #1 \earr\right]}
\def\Bf{\bm{B}}
\def\gbh{\widehat{\gb}}
\def\dbh{\widehat{\db}}
\def\Tr#1{\mbox{Tr}\left[#1\right]}
\def\Cov#1{\mbox{Cov}\left[#1\right]}
\def\iid{i.i.d.~}

\def\cro#1{\left[#1\right]}
\def\acc#1{\left\{#1\right\}}
\def\aprio{\emph{a priori~}}
\def\apost{\emph{a posteriori~}}
\def\hbh{\widehat{\hb}}
\def\qh{\widehat{q}}
\def\fbt{\widetilde{\fb}}
\def\thetabt{\widetilde{\thetab}}



\def\bm#1{\mbox{\boldmath $#1$}}

\def\zerob{{\bm 0}}
\def\oneb{{\bm 1}}

\def\ab{{\bm a}}
\def\bb{{\bm b}}
\def\cb{{\bm c}}
\def\db{{\bm d}}
\def\eb{{\bm e}}
\def\fb{{\bm f}}
\def\gb{{\bm g}}
\def\hb{{\bm h}}
\def\ib{{\bm i}}
\def\jb{{\bm j}}
\def\kb{{\bm k}}
\def\lb{{\bm l}}
\def\mb{{\bm m}}
\def\nb{{\bm n}}
\def\ob{{\bm o}}
\def\pb{{\bm p}}
\def\qb{{\bm q}}
\def\rb{{\bm r}}
\def\sb{{\bm s}}
\def\tb{{\bm t}}
\def\ub{{\bm u}}
\def\vb{{\bm v}}
\def\wb{{\bm w}}
\def\xb{{\bm x}}
\def\yb{{\bm y}}
\def\zb{{\bm z}}

\def\Ab{{\bm A}}
\def\Bb{{\bm B}}
\def\Cb{{\bm C}}
\def\Db{{\bm D}}
\def\Eb{{\bm E}}
\def\Fb{{\bm F}}
\def\Gb{{\bm G}}
\def\Hb{{\bm H}}
\def\Ib{{\bm I}}
\def\Jb{{\bm J}}
\def\Kb{{\bm K}}
\def\Lb{{\bm L}}
\def\Mb{{\bm M}}
\def\Nb{{\bm N}}
\def\Ob{{\bm O}}
\def\Pb{{\bm P}}
\def\Qb{{\bm Q}}
\def\Rb{{\bm R}}
\def\Sb{{\bm S}}
\def\Tb{{\bm T}}
\def\Ub{{\bm U}}
\def\Vb{{\bm V}}
\def\Wb{{\bm W}}
\def\Xb{{\bm X}}
\def\Yb{{\bm Y}}
\def\Zb{{\bm Z}}

\def\alphab{\bm{\alpha}}
\def\betab{\bm{\beta}}
\def\deltab{\bm{\delta}}
\def\epsilonb{\bm{\epsilon}}
\def\gammab{\bm{\gamma}}
\def\omegab{\bm{\omega}}
\def\thetab{\bm{\theta}}
\def\xib{\bm{\xi}}
\def\lambdab{\bm{\lambda}}
\def\taub{\bm{\tau}}
\def\phib{\bm{\phi}}
\def\mub{\bm{\mu}}
\def\psib{\bm{\psi}}
\def\chib{\bm{\chi}}
\def\sigmab{\bm{\sigma}}

\def\Deltab{\bm{\Delta}}
\def\Lambdab{\bm{\Lambda}}
\def\Phib{\bm{\Phi}}
\def\Psib{\bm{\Psi}}
\def\Sigmab{\bm{\Sigma}}

\def\Ac{{\cal A}}
\def\Bc{{\cal B}}
\def\Cc{{\cal C}}
\def\Dc{{\cal D}}
\def\Ec{{\cal E}}
\def\Fc{{\cal F}}
\def\Gc{{\cal G}}
\def\Hc{{\cal H}}
\def\Ic{{\cal I}}
\def\Jc{{\cal J}}
\def\Kc{{\cal K}}
\def\Lc{{\cal L}}
\def\Mc{{\cal M}}
\def\Nc{{\cal N}}
\def\Oc{{\cal O}}
\def\Pc{{\cal P}}
\def\Qc{{\cal Q}}
\def\Rc{{\cal R}}
\def\Sc{{\cal S}}
\def\Tc{{\cal T}}
\def\Uc{{\cal U}}
\def\Vc{{\cal V}}
\def\Wc{{\cal W}}
\def\Xc{{\cal X}}
\def\Yc{{\cal Y}}
\def\Zc{{\cal Z}}

\def\wt#1{\widetilde{#1}}
\def\wh#1{\widehat{#1}}
%

\def\xh{\widehat{x}}
\def\thetah{\widehat{\theta}}
\def\betah{\widehat{\beta}}

\def\xbh{\widehat{\xb}}
\def\thetabh{\widehat{\thetab}}
\def\betabh{\widehat{\betab}}

\def\xbhk{\widehat{\xb}^{k}}
\def\thetahk{\widehat{\theta}^{k}}
\def\betahk{\widehat{\beta}^{k}}

\def\xbhkp{\widehat{\xb}^{k+1}}
\def\thetahkp{\widehat{\theta}^{k+1}}
\def\betahkp{\widehat{\beta}^{k+1}}

\def\thetabhk{\widehat{\thetab}^{k}}
\def\betabhk{\widehat{\betab}^{k}}

\def\thetabhkp{\widehat{\thetab}^{k+1}}
\def\betabhkp{\widehat{\betab}^{k+1}}

\def\thetamin{\theta_{\mbox{\tiny min}}}
\def\thetamax{\theta_{\mbox{\tiny max}}}
\def\betamin{\beta_{\mbox{\tiny min}}}
\def\betamax{\beta_{\mbox{\tiny max}}}

\def\ra{\rightarrow}
\def\la{\leftarrow}
\def\da{\downarrow}
\def\ua{\uparrow}

\def\Ra{\Rightarrow}
\def\La{\Leftarrow}
\def\Da{\Downarrow}
\def\Ua{\Uparrow}

\def\lra{\longrightarrow}
\def\lla{\longleftarrow}
\def\Lra{\Longrightarrow}
\def\Lla{\longleftarrow}

\def\lrarr{\leftrightarrow}
\def\Lrarr{\Leftrightarrow}
\def\udarr{\updownarrow}
\def\Uparr{\Updownarrow}

\def\d#1{\,\mbox{d}#1}
\def\dxdy{\d{x}\d{y}}
\def\dwxdwy{\d{\omega_x}\d{\omega_y}}
\def\dxdydz{\d{x}\d{y}\d{z}}

\def\disp#1{{\displaystyle #1}}
\def\diag#1{\mbox{diag}\left\{#1\right\}}

\def\Prob#1{\mbox{Pr}\left\{#1\right\}}
\def\var#1{\mbox{Var}\left\{#1\right\}}
\def\cov#1{\mbox{Cov}\left\{#1\right\}}
\def\corr#1{\mbox{Corr}\left\{#1\right\}}
\def\trace#1{\mbox{Tr}\left\{#1\right\}}
\def\rang#1{\mbox{rang}\left\{#1\right\}}
\def\det#1{\mbox{d\'et}\left\{#1\right\}}

\def\cosf{\cos \phi}
\def\sinf{\sin \phi}
\def\cost{\cos \theta}
\def\sint{\sin \theta}

\def\sgn{\mbox{sgn}}
\def\sinc{\mbox{sinc}}
\def\rect{\mbox{rect}}
\def\sincf#1{\mbox{sinc}\left(#1\right)}
\def\rectf#1{\mbox{rect}\left(#1\right)}
\def\trif#1{\mbox{tri}\left(#1\right)}
%
\def\xvec#1#2#3{\left\{#1_#2,\ldots,#1_#3\right\}}

\def\vx{\left[x_1,\ldots, x_n\right]^t}
\def\vz{\left[z_1,\ldots, z_n\right]^t}
\def\vw{\left[\omega_1,\ldots, \omega_n\right]^t}
\def\vxi{\left[\xi_1,\ldots, \xi_n\right]^t}
%
\def\iii{\int_{-\infty}^{+\infty}}
\def\izi{\int_{0}^{\infty}}
\def\izpi{\int_{0}^{\pi}}
\def\izdpi{\int_{0}^{2\pi}}
\def\intd{\int\kern-.8em\int}
\def\intt{\int\kern-.8em\int\kern-.8em\int}
\def\intg{\int\kern-1.1em\int}
\def\sumd{\mathop{\sum\sum}}

\def\sumi{\sum_{i=1}^{M}}
\def\sumj{\sum_{i=1}^{N}}
\def\sumk{\sum_{k=1}^{K}}
\def\sumn{\sum_{n=1}^{N}}
\def\summ{\sum_{m=1}^{M}}

%
\def\TA#1{{\cal A}\left\{ {#1} \right\}}
\def\TH#1{{\cal H}\left\{ {#1} \right\}}
\def\TP#1{{\cal P}\left\{ {#1} \right\}}
\def\TR#1{{\cal R}\left\{ {#1} \right\}}
\def\TRa#1{{\cal R}^{\dag}\left\{ {#1} \right\}}
\def\BR#1{{\cal B}\left\{ {#1} \right\}}
\def\TF#1{{\cal F}\left\{ {#1} \right\}}
\def\TFI#1{{\cal F}^{-1}\left\{ {#1} \right\}}
\def\TFn#1#2{{\cal F}_{#1}\left\{ {#2} \right\}}
\def\TFnI#1#2{{\cal F}_{#1}^{-1}\left\{ {#2} \right\}}
\def\Im#1{{\cal I}\mbox{m}\left(#1\right)}
\def\Ker#1{{\cal K}\mbox{er}\left(#1\right)}
\def\Imag#1{\mbox{Im}\left(#1\right)}
\def\Re#1{\mbox{Re}\left(#1\right)}
\def\expf#1{\exp\left[ {#1} \right]}

\def\dfdx#1#2{{\mbox{d} {#1}\over{\mbox{d} {#2}}}}
\def\dfdxd#1#2{{\mbox{d}^2 {#1}\over{\mbox{d} {#2}^2}}}
\def\dfdxt#1#2{{\mbox{d}^3 {#1}\over{\mbox{d} {#2}^3}}}
\def\dfdxn#1#2{{\mbox{d}^n {#1}\over{\mbox{d} {#2}^n}}}
\def\dfdxk#1#2{{\mbox{d}^k {#1}\over{\mbox{d} {#2}^k}}}

\def\dpdx#1#2{\frac{\partial {#1}}{\partial {#2}}}
\def\dpdxd#1#2{\frac{\partial^2 {#1}}{\partial {#2}^2}}
\def\dpdxdy#1#2#3{{{\partial ^2 {#1}}\over{\partial {#2} \partial {#3}}}}

\def\arg{\mbox{arg}}
\def\argmins#1#2{\mbox{arg}\min_{#1}\left\{{#2}\right\}}
\def\argmaxs#1#2{\mbox{arg}\max_{#1}\left\{{#2}\right\}}
\def\argmin#1#2{\mathop{\mbox{arg}\min}_{#1}\left\{{#2}\right\}}
\def\argmax#1#2{\mathop{\mbox{arg}\max}_{#1}\left\{{#2}\right\}}

\def\esp#1{\mbox{E}\left\{ #1 \right\}}
\def\espx#1#2{\mbox{E}_{#1}\left\{ #2 \right\}}

\def\wth#1{\widehat{\widetilde{\phantom{#1}}}\!\!\!\! #1}

\def\lrf{L_{r,\phi}}
\def\fw{\widehat{f}(\omegab)}
\def\fthwxi{\wth{f}(\Omega,\xib)}
\def\fthwfi{\wth{f}(\Omega,\phi)}
\def\ftrfi{\widetilde{f}(r,\phi)}

\def\fwxwy{\widehat{f}(\omega_x, \omega_y)}
\def\wxpwy{(\omega_x \, x + \omega_y \, y)}

\def\wtx{\omegab^t \cdot \xb}
\def\ejwtx{\exp\left[j \omegab^t \cdot \xb\right]}
\def\xitx{\xib^t \cdot \xb}

\def\ftrxi{\widetilde{f}(r,\xib)}
\def\ent{-\int p(x) \, \ln p(x) \d{x}}


\def\mean#1{\left< #1 \right>}
\def\slnhn{\sum_{n=1}^N \lambda_n h_n(\rb)}
\def\slngn{\sum_{n=1}^N \lambda_n g_n(\rb)}
\def\smngn{\sum_{n=1}^N \mu_n g_n(\rb)}
\def\slmhm{\sum_{m=1}^N \lambda_m h_m(\rb)}
\def\vlambda{\bm{\lambda} = [\lambda_1,\ldots,\lambda_n]}

\def\apriori{{\em a priori} }
\def\aposteriori{{\em a posteriori} }

\def\titre#1{\bcc{\Large\bf #1}\ecc}

\def\AMD{Ali Mohammad--Djafari}
\def\LSSa{Laboratoire des Signaux et Syst\`emes 
(CNRS--ESE--UPS) \\ 
\'Ecole Sup\'erieure d'\'Electricit\'e \\ 
Plateau de Moulon, 91192 Gif sur Yvette Cedex, France.}

\def\ME{maximum entropy}
\def\pdf{probability distribution function}
\def\lm{Lagrange multipliers}
\def\fix#1{\phi _#1(x)}
\def\fin{\fix n}
\def\fik{\fix k}
\def\fiz{\fix 0}
\def\sfinz{\sum_{n=0}^N \lambda_n \, \fin}
\def\sfinu{\sum_{n=1}^N \lambda_n \, \fin}
\def\bl{\bm{\lambda}}
\def\bd{\bm{\delta}}
\def\blz{\bl ^0}
\def\gnl{G _n(\bl)}
\def\gnlz{G _n(\blz)}
\def\un{n=1,\dots, N}
\def\nn{n=0,\dots, N}

\def\finn{\fin , \nn}
\def\esfinz{\exp\,\left[ -\sfinz \right] }
\def\esfinu{\exp\,\left[ -\sfinu \right] }
\def\esxm{\exp\,\left[ -\sum_{m=0}^N \lambda_m \, x^m \right] }
\def\efin{\esp \fin }
\def\zl{Z(\bl)}
\def\finxi{\phi _n(x_i)}
\def\snfinxi{\sum_{n=1}^N \lambda_n \finxi}
\def\esnfinxi{\exp \left[ - \snfinxi \right]}
\def\smfinxi{\sum_{i=1}^M \finxi}

\def\ejnw{\exp \left( -j n \omega_0 x \right) }
\def\eejnw{\mbox{E} \left\lbrace \ejnw \right\rbrace}

\def\signed#1{{\unskip\nobreak\hfil\penalty50\hskip2em\mbox{}
\nobreak\hfil\tt#1\parfillskip=0pt \finalhyphendemerits=0 \par}}

\def\uncatcodespecials{\def\do##1{\catcode`##1=12 }\dospecials}
\def\listing#1{\par\begingroup\setupverbatim\input#1 \endgroup}
\newcount\lineno
\def\setupverbatim{\tt \lineno=0
 \obeylines \uncatcodespecials \obeyspaces
 \everypar{\advance\lineno by1 \llap{\sevenrm\the\lineno\ \ }}}
{\obeyspaces\global\let =\ }

\def\defined{\stackrel{\mbox{def}}{=}}
\def\str{\stackrel}

\def\ER{\mbox{I\kern-.25em R}}
\def\EC{\mbox{C\kern-.8em C}}
\def\EZ{\mbox{Z\kern-.55em Z}}
\def\EN{\mbox{N\kern-.8em N}}

\def\singles{
 \abovedisplayskip 12pt plus 3pt minus 9pt
 \belowdisplayskip 12pt plus 3pt minus 9pt
 \abovedisplayshortskip 0pt plus 3pt
 \belowdisplayshortskip 7pt plus 3pt minus 4pt
 \baselineskip 14.4pt
 \lineskip 1pt
 \lineskiplimit 0pt}

\def\oneandhalf{
 \abovedisplayskip 18pt plus 3pt minus 9pt
 \belowdisplayskip 18pt plus 3pt minus 9pt
 \abovedisplayshortskip 0pt plus 3pt
 \belowdisplayshortskip 9.333pt plus 3pt
 \baselineskip 20pt
 \lineskip 2pt
 \lineskiplimit 1pt}

\def\double{
 \abovedisplayskip 24pt plus 3pt minus 9pt
 \belowdisplayskip 24pt plus 3pt minus 9pt
 \abovedisplayshortskip 0pt plus 3pt
 \belowdisplayshortskip 12pt plus 3pt
 \baselineskip 27pt
 \lineskip 3pt
 \lineskiplimit 2pt}

\def\dadb{\d{\alpha}\d{\beta}}

\def\ffbox#1{\fbox{\mbox{\vbox{#1}}}}

\def\rot{\mbox{rot}}
\def\case#1#2#3#4{
    \left\{
           \begin{array}{ll}
            {\displaystyle #1} & {\displaystyle #2} \cr 
            {\displaystyle #3} & {\displaystyle #4}
           \end{array}
    \right. }

\def\beqnarr#1&#2&#3\\#4&#5&#6\eeqnarr{
    \left\{
           \begin{array}{lcl}
            {\displaystyle #1} & #2 & {\displaystyle #3} \\ 
            {\displaystyle #4} & #5 & {\displaystyle #6} 
           \end{array}
    \right. }

\def\pyx{p(\yb|\xb)}
\def\pxy{p(\xb|\yb)}

\def\ie{{\em i.e.}}
\def\unsdpi{\left(\frac{1}{2\pi}\right)}
\def\unspi{\left(\frac{1}{\pi}\right)}
\def\up{\uppercase}
\def\zjm{z_{j-1}}
\def\zjp{z_{j+1}}
\def\fxyp{f(x,y)=\left\{
\barr{ll} 1 & (x,y)\in P\\ 0 & (x,y)\not\in P\earr
\right.}

\maketitle

\section{Introduction}
Dans une approche estimation bay\'esienne non supervis\'ee pour r\'esoudre un probl\`eme inverse, on commence par \'ecrire l'expression de la loi \apost conjointe des inconnues $\fb$ et des hyper-param\`etres $\thetab$~:
\beq
p(\fb,\thetab|\gb;\Mc)=\frac{p(\gb,\fb,\thetab|\Mc)}{p(\gb|\Mc)}
=\frac{p(\gb|\fb,\thetab;\Mc) \; p(\fb,\thetab|\Mc)}{p(\gb|\Mc)}.
\eeq
Dans cette relation $p(\gb|\fb,\thetab;\Mc)$ est la vraisemblance des inconnues dont l'expression s'obtient \`a partir d'un mod\`ele liant les inconnues aux donn\'ees $\gb$ (mod\'elisation du probl\`eme directe), $p(\fb,\thetab|\Mc)$ est la loi \aprio des inconnues et 
\beq
p(\gb|\Mc)= \intd p(\gb|\fb,\thetab;\Mc) \; p(\fb|\thetab;\Mc) \; p(\thetab|\Mc) \d{\fb} \d{\thetab}
\eeq
est ce qu'on appelle l'\'evidence du mod\`ele $\Mc$. 

Il est int\'eressant de mentionner que, pour n'importe quelle loi de 
probabilit\'e $q(\fb,\thetab)$ on a
\beqn
p(\gb|\Mc)
&=& \intd p(\gb,\fb,\thetab|\Mc) \d{\fb} \d{\thetab} \nonumber \\ 
&=& \intd q(\fb,\thetab) \frac{p(\gb,\fb,\thetab|\Mc)}{q(\fb,\thetab)} \d{\fb} \d{\thetab} \nonumber \\ 
&\ge& \intd q(\fb,\thetab) \ln \frac{p(\gb,\fb,\thetab|\Mc)}{q(\fb,\thetab)} \d{\fb} \d{\thetab}.
\eeqn
Aussi, notant par  
\vspace{-9pt}
\beq
\Fc(q)=\intd q(\fb,\thetab) \ln \frac{p(\gb,\fb,\thetab|\Mc)}{q(\fb,\thetab)} \d{\fb} \d{\thetab}
\eeq

\vspace{-12pt}
\noindent et par 
\vspace{-12pt}
\beq
\mbox{KL}(q:p)=\intd q(\fb,\thetab) \ln \frac{p(\fb,\thetab|\gb;\Mc)}{q(\fb,\thetab)} \d{\fb} \d{\thetab}
\eeq 
\vspace{-12pt}
on montre facilement que

\beq
p(\gb|\Mc)=\Fc(q)+\mbox{KL}(q:p).
\eeq 
Ainsi $\Fc(q)$, appel\'ee l'\'energie libre de $q$ par rapport \`a $p$, est une limite inf\'erieure de 
$p(\gb|\Mc)$ car $\mbox{KL}(q:p)\ge 0$. 
Par la suite, nous allons \'ecrire l'expression de $\Fc(q)$ par 
\beq \label{eq4}
\Fc(q)= \left<\ln{p(\gb,\fb,\thetab;\Mc)}\right>_q+\Hc(q)
\eeq
o\`u $\Hc(q)$ est l'entropie de $q$. 
\cite{Ghahramani97,Choudrey03a}

\section{Approche variationnelle}
Nous allons maintenant utiliser ces relation pour d\'ecrire le principe de l'approche variationnelle. 
L'id\'ee de base est que l'utilisation directe de la loi \apost conjointe \\ 
$p(\fb,\thetab|\gb;\Mc)$ 
est souvent tr\`es co\^uteux pour, par exemple, \^etre explor\'ee par \'echantillonnage directe 
ou pour calculer les moyennes \apost 
\hfil $\fbh=\int\int \fb \; {p(\fb,\thetab|\gb;\Mc)} \d{\thetab} \d{\fb}$ \\ 
\mbox{~et~} 
$\thetabh=\int\int \; \thetab {p(\fb,\thetab|\gb;\Mc)} \d{\fb} \d{\thetab}$. 
En effet, rare sont les cas o\`u on puisse trouver des expressions analytiques pour ces int\'egrales. 
De m\^eme l'exploration de cette loi par des m\'ethodes de Mont\'e Carlo est aussi co\^uteuses. 
On cherche alors de l'approximer 
par une loi plus simple $q(\fb,\thetab)$. Par simplicit\'e, nous entendons par exemple une loi $q$ qui soit s\'eparable en $\fb$ et en $\thetab$~:
\beq
q(\fb,\thetab)= q_1(\fb) \; q_2(\thetab)
\eeq
\'Evidemment, cette approximation doit \^etre fait de telle sorte qu'une mesure de distance entre $q$ et $p$ soit minimale. Si, d'une mani\`ere naturelle, on choisi $\mbox{KL}(q:p)$ comme cette mesure, on aura:
\beq
(\qh_1, \qh_2)=\argmin{(q_1, q_2)}{\mbox{KL}(q_1 q_2 : p)}=\argmax{(q_1, q_2)}{\Fc(q_1 q_2)}
\eeq
et sachant que $\mbox{KL}(q_1 q_2 : p)$ est convexe en $q_1$ \`a , $q_2$ fix\'ee et vise versa, 
on peut obtenir la solution d'une mani\`ere it\'erative~: 
\beq
\left\{\barr{ll}
\qh_1&=\argmin{q_1}{\mbox{KL}(q_1 \qh_2 : p)}=\argmax{q_1}{\Fc(q_1 \qh_2)} \\ 
\qh_2&=\argmin{q_2}{\mbox{KL}(\qh_1 q_2 : p)}=\argmax{q_2}{\Fc(\qh_1 q_2)}
\earr\right.
\eeq
Utilisant la relation (\ref{eq4}), il est facile de montrer que les solutions d'optimisation de  
de ces \'etapes sont 
\beq
\left\{\barr{ll}
\qh_1(\fb)&\propto \expf{\left<\ln{p(\gb,\fb,\thetab;\Mc)}\right>_{\qh_2(\thetab)}} \\ 
\qh_2(\thetab)&\propto \expf{\left<\ln{p(\gb,\fb,\thetab;\Mc)}\right>_{\qh_1(\fb)}} 
\earr\right.
\eeq
Une fois cet algorithme converg\'e vers $\qh^*_1(\fb)$ et $\qh^*_2(\thetab)$, on peut les utiliser 
d'une mani\`ere ind\'ependante pour calculer, par exemple les moyennes \quad 
$\fbh^*    = \int \fb \; \qh^*_1(\fb) \d{\fb}$  \quad et \\ 
$\thetabh^*= \int \thetab \;\; \qh^*_2(\thetab) \d{\thetab}$.

Une deuxi\`eme \'etape de simplification est n\'ecessaire pour \^etre capable de calculer les esp\'erances 
qui se trouvent dans ces exponentielles. Les calculs non param\'etriques sont souvent trop co\^uteux. 
On choisit alors une forme param\'etrique pour ces lois de telle sorte qu'on puisse, \`a chaque it\'eration, 
remettre \`a jours seulement les param\`etres de ces lois, \`a condition cependant que ces formes ne 
changent pas au cours des it\'erations. La famille des lois exponentielles conjugu\'ees ont cette propri\'et\'e  
\cite{Ghahramani97,Patriksson99,Choudrey03b,Molina99,vb_gar_ieee,Nasios2006,Likas2004}. 
Nous examinons ici, trois cas:

\subsection{Cas d\'eg\'en\'er\'ee} 

Il s'agit de choisir pour $\qh_1(\fb)$ et $\qh_2(\thetab)$ des formes d\'eg\'en\'er\'ees suivantes~:
\beq
\left\{\barr{ll}
\qh_1(\fb|\fbt)&=\delta(\fb-\fbt) \\ 
\qh_2(\thetab|\thetabt)&=\delta(\thetab-\thetabt) 
\earr\right.
\eeq
Par cons\'equence, qu'au cours des it\'erations, nous aurons \`a remettre 
\`a jours $\fbt$ et  $\thetabt$ au cours des it\'erations.  
 les param\`etres de la loi \apost jointe $p(\fb,\thetab|\gb;\Mc)$. 

En remarquant alors que 
\beq
\left\{\barr{ll}
\qh_1(\fb)\propto p(\gb,\fb,\thetabt;\Mc) \propto p(\fb,\thetabt|\gb;\Mc)  \\ 
\qh_2(\thetab) \propto p(\gb,\fbt,\thetab;\Mc) \propto p(\fbt,\thetab|\gb;\Mc)
\earr\right.
\eeq
Il est alors facile de voir que si $p(\fb,\thetab|\gb;\Mc)$ est gaussienne \`a $\thetab$ fix\'e, 
on aura juste \`a calculer 
$\fbt=\argmax{\fb}{p(\fb,\thetabt|\gb;\Mc)}$ que l'on utilise ensuite pour mettre \`a jour 
$\qh_2(\thetab)$. On note alors que cet algorithme devient \'equivalent \`a ce qu'on peut 
apeller MAP Joint~:
\beq
\left\{\barr{@{}l@{}l@{}l@{}}
\fbt    &=\argmax{\fb}{p(\gb,\fb,\thetabt;\Mc)}    &=\argmax{\fb}{p(\fb,\thetabt|\gb;\Mc)} \\ 
\thetabt&=\argmax{\thetab}{p(\gb,\fbt,\thetab;\Mc)}&=\argmax{\fb}{p(\fbt,\thetab|\gb;\Mc)}   
\earr\right.
\eeq
On remarque que l'on retrouve un algorithme du type MAP jointe. 

\subsection{Cas particulier conduisant \`a l'algorithme EM} 
Il s'agit de choisir, comme dans le cas pr\'ec\'edent une forme d\'eg\'en\'er\'ee pour 
$\qh_2(\thetab)=\delta(\thetab-\thetabt)$, ce qui donne 
\beq
\qh_1(\fb)\propto p(\gb,\fb,\thetabt;\Mc) \propto p(\fb,\thetabt|\gb;\Mc)\propto p(\fb|\thetabt,\gb;\Mc)
\eeq
ce qui signifie que $\qh_1(\fb)$ est une loi dans la m\^eme famille que la loi \apost 
$p(\fb|\gb,\thetab;\Mc)$. \'Evidemment, si la forme de cette loi est simple, par exemple une gaussienne,  
(ce qui est le cas dans les situations que nous \'etudierons) les calculs seront simples. 

A chaque it\'eration, on aurait alors \`a remettre \`a jours $\thetabt$ 
qui est ensuite utilis\'e pour trouver $\qh_1(\fb|\thetabt)=p(\fb|\gb,\thetabt;\Mc)$, qui est utilis\'ee 
pour calculer 
\beq
Q(\thetab,\thetabt)=\left<\ln{p(\gb,\fb,\thetab;\Mc)}\right>_{\qh_1(\fb|\thetabt)}\eeq
On remarque facilement l'\'equivalence avec l'algorithme EM. 

\subsection{Cas particulier propos\'e pour les probl\`emes inverses} 

Il s'agit de choisir, pour $\qh_1(\fb)$ et $\qh_2(\thetab)$ les m\^emes familles de lois que 
$p(\fb|\gb,\thetab)$ et $p(\thetab|\gb,\fb)$, ce qui permet de profiter de la mise \`a jour facile 
de ces lois si des lois \aprio correspondantes sont choisie dans les familles des lois conjugu\'ees 
associ\'e \`a la mod\'elisation directe du probl\`eme.  

Dans ce travail, dans un premier temps, nous allons consid\'er\'e le cas des 
probl\`emes inverses lin\'eaires~:
\beq 
\gb=\Hb \; \fb + \epsilonb 
\label{eq2}
\eeq
o\`u $\Hb$ repr\'esente la forme discr\'etis\'e de la mod\'elisation directe du probl\`eme et 
$\epsilonb$ repr\'esente l'ensemble des erreurs de mesure et de mod\'elisation  avec des hypoth\`eses suivantes:   
\beq
\barr{r@{}c@{}l}
p(\gb|\Hb,\fb,\theta_e;\Mc)     &=& \Nc(\Hb\fb,(1/\theta_e)\Ib), \\ 
p(\fb|\theta_f;\Mc)             &=& \Nc(\zerob,(1/\theta_f)(\Db^t_f\Db_f)^{-1}),\\ 
p(\theta_e;\Mc)                  &=& \Gc(\alpha_{e0},\beta_{e0}),\\ 
p(\theta_f;\Mc)                  &=& \Gc(\alpha_{f0},\beta_{f0})
\earr 
\eeq
o\`u $\thetab=(\theta_e=1/\sigma_{\epsilon}^2,\theta_f=1/\sigma_f^2)$. 
On obtient alors facilement les expressions de $p(\gb,\fb|\thetab;\Mc)$, $p(\fb|\gb,\thetab;\Mc)$ et  $p(\thetab|\gb,\fb;\Mc)$ qui sont~:
\beq
\barr{r@{}c@{}l}
p(\gb,\fb|\Hb,\theta_e;\Mc) &=& \Nc(\Hb\fb,(1/\theta_e)\Ib)\; \Nc(\zerob,(1/\theta_f)(\Db^t_f\Db_f)^{-1}), \\ 
p(\fb|\gb,\Hb,\theta_f;\Mc) &=& \Nc(\fbh,\Sigmabh),\\ 
p(\theta_e;\Mc)                  &=& \Gc(\alphah_{e},\betah_{e}), \\ 
p(\theta_f;\Mc)                   &=& \Gc(\alphah_{f},\betah_{f})
\earr
\eeq
o\`u les expressions de $\fbh$, $\Sigmabh$, $(\alphah_{e},\betah_{e})$ et 
$(\alphah_{f},\betah_{f})$ seront donn\'ees en annexe. 

\section{Application en restauration d'image}
Dans le cas de la restauration d'image o\`u $\Hb$ a une structure particuli\`ere, et o\`u l'op\'eration 
$\Hb\fb$ repr\'esente une convolution de l'image $f$ avec la r\'eponse impulsionnelle $h$, la partie difficile et co\^uteuse de ces calculs est celle du calcul de $\fbh$ qui peut se faire \`a l'aide de la Transform\'ee de Fourier rapide. 

De m\^eme, l'approche peut tr\`es facilement \^etre \'etendue pour le cas de la 
restauration aveugle ou myope o\`u on cherche \`a la fois d'estimer la r\'eponse pulsionnelle $\hb$, 
l'image $\fb$ et les hyper-param\`etres $\thetab$. 
Pour \'etablir l'expressions des diff\'erentes lois dans ce cas, nous notons que le probl\`eme directe, suivant que l'on s'int\'eresse \`a $\fb$ (d\'econvolution) ou  \`a $\hb$ (identification de la r\'eponse impulsionnelle), peux 
s'\'ecrire 
\beq 
\barr{r@{}c@{}l}
g(\rb)&=&h(\rb)*f(\rb)+\epsilon(\rb) =f(\rb)*h(\rb)+\epsilon(\rb) \\ 
\gb&=&\Hb \; \fb + \epsilonb =\Fb \; \hb + \epsilonb
\earr
\eeq
Pour permettre d'obtenir une solution bay\'esienne pour l'\'etape de l'identification, nous devons 
aussi mod\'eliser $\hb$. Une solution est de supposer $\hb=\Phib \wb$ o\`u la matrice $\Phib$ 
est une matrice telle que $\Phib \wb$ repr\'esente la convolution $\phi(\rb)*w(\rb)$. 
Ainsi les colonnes de $\Phib$ repr\'esentent une base et les \'el\'ements du vecteur $\wb$ repr\'esentent 
les co\'efficients de la d\'ecomposition de $h$ sur cette base. On a ainsi
\beq 
\barr{r@{}c@{}l}
g(\rb)&=&(\phi*w)*f(\rb)+\epsilon(\rb) =f*(\phi*w)(\rb)+\epsilon(\rb) \\ 
\gb&=&\Phib \; \Wb \; \fb + \epsilonb =\Fb \; \Phib \; \wb + \epsilonb
\earr
\eeq
Le probl\`eme de la d\'econvolution aveugle se ram\`ene \`a l'estimation de $\fb$ et $\wb$ 
avec des lois
\beq
\barr{c}
p(\gb|\wb,\fb,\Sigmabe) = \Nc(\Phib\Wb\fb,\Sigmabe)
= \Nc(\Fb\Phib\wb,\Sigmabe), \\ 
\mbox{~avec~} \Sigmabe=\diag{\frac{1}{\theta_{ei}}, i=1,\cdots,M} 
\mbox{~et~} p(\theta_{ei})= \Gc(\alpha_{e0},\beta_{e0}) 
\\ 
p(\fb|\theta_f)             = \Nc\left(\zerob,(\theta_f\Db^t_f\Db_f)^{-1}\right)  
\mbox{~avec~} p(\theta_f)= \Gc(\alpha_{f0},\beta_{f0}),\\ 
p(\wb|\alphab)            = \prod_j \Nc(0,\frac{1}{\alpha_j}) 
\mbox{~avec~}p(\alphab) = \prod_j \Gc(a_0,b_0), \forall j 
\earr 
\eeq
Avec ces lois \aprio, il est alors facile de trouver l'expression de la loi conjointe 
$p(\fb,\wb,\thetab_e,\theta_f,\alphab;\gb)$ et la loi \apost  $p(\fb,\wb,\thetab_e,\theta_f,\alphab|\gb)$. 
Cependant l'expression de cette loi  
\beq
\barr{ll}
p(\fb,\wb,\thetab_e,\theta_f,\alphab|\gb) \propto & 
p(\gb|\wb,\fb,\Sigmabe) \,  p(\fb|\theta_f) \, p(\wb|\alphab) \\ 
& p(\theta_e) \, p(\theta_f) \, p(\alphab)
\earr 
\eeq
n'est pas s\'eparable en ses composantes. L'approche variationnelle consiste donc \`a l'approximer par une loi s\'eparable \\ 
\hfil\(
p(\fb,\wb,\thetab_e,\theta_f,\alphab|\gb) \simeq q(\fb) q(\wb) \prod_j q(\theta_{ei}) q(\theta_f) \prod_j q(\alpha_j)
\)
et avec les choix des lois \aprio conjugu\'ees en appliquant la proc\'edure d\'ecrite plus haut, on obtient 
\beqn
q(\fb)    &=& \Nc(\mub_f,\Sigmab_f) \mbox{~avec~} \nonumber\\ 
\Sigmab_f &=& [\Phib^t <\Wb^t<\Bf> \Wb> \Phib+<\theta_f>\Qb^t\Qb]^{-1}, \nonumber\\ 
\mub_f    &=& \Sigmab_f \Phib^t  <\Wb>^t <\Bf> \gb,  
\\
q(\wb)    &=& \Nc(\mub_w,\Sigmab_w) \mbox{~avec~} \nonumber\\  
\Sigmab_w &=& [\Phib^t <\Fb^t<\Bf> \Fb>\Phib+\Ab]^{-1}, \nonumber\\ 
\mub_w    &=& \Sigmab_w \Phib^t  <\Fb>^t <\Bf> \gb 
\\ 
q(\theta_{ei})         &=& \Gc(\alpha_{ei},\beta_{ei}) \mbox{~avec~} \nonumber\\  
\alpha_{ei}                &=& \alpha_{e0} + M/2, \nonumber\\ 
\beta_{ei}                &=& \beta_{e0} + 1/2 <\epsilonb\epsilonb^t>_{ii}, 
\\
q(\theta_f)            &=& \Gc(\alpha_{f},\beta_{f}) \mbox{~avec~} \nonumber\\ 
\alpha_{f}                &=& \alpha_{f0} + N/2,  \nonumber\\
\beta_{f}                &=& \beta_{f0} + 1/2 \, \trace{\Qb^t\Qb<\fb\fb^t>},  
\\
q(\alpha_{wj})         &=& \Gc(a_{j},b_{j}) \mbox{~avec~} \nonumber\\ 
\alpha_{wj}                &=& \alpha_{w0} + 1/2, \nonumber \\
\beta_{w}                &=& \beta_{w0} + 1/2 <w_i^2>.
\eeqn
o\`u $\Ab=\diag{\alpha_{wj}, j=1,\cdots,N}$ et $\Bb=\diag{\beta_{ei}, i=1,\cdots,M}$. 

On a ainsi l'expression des diff\'erentes composante de la loi s\'eparable approchante. On peut en d\'eduire facilement les moyennes des ces lois, car ces lois sont, soit des gaussiennes, soit des lois gamma. 
\beq
\barr{r@{}c@{}l}
<\wb> &=& \mub_w, \quad <w_j^2>=[\mub_w]_i^2+[\Sigmab_w]_{jj}\\ 
<\fb> &=& \mub_f, \quad <\fb\fb^t>=\mub_f\mub_f^t + \Sigmab_f\\ 
<\alpha_{ei}>                &=& \alpha_{ei}/\beta_{ei}, \quad 
<\beta_{ei}>                = \alpha_{ei}/\beta_{ei}^2, \\
<\alpha_{f}>                &=& \alpha_{f}/\beta_{f}, \quad
<\beta_{f}>                = \alpha_{f}/\beta_{f}^2, \\
\earr
\eeq
\beq
<\epsilonb\epsilonb^t>=\gb\gb^t-2\gb [<\Fb> \Phib <\wb>]^t+\Phib [\Fb \wb\wb^t \Fb^t] \Phib^t
\eeq
Pour le calcul des termes $<\Wb^t\Wb>$ et $<\Fb^t\Fb>$ qui interviennent dans les expressions de $\Sigmab_f$, $\Sigmab_w$ et $[\Fb \wb\wb^t \Fb^t]$ on peut utiliser le fait que $\Fb$ et $\Wb$ sont des matrices block-Toeplitz avec des blocs Toeplitz (TBT), on peut les approximer par des matrices block-circulantes avec des blocs circulantes (CBC) et les inverser en utilisant la TFD. 
Notons aussi que $\mub_f$ et $\mub_w$ peuvent \^etre obtenu par optimisation de 
\beq
\barr{r@{}c@{}l}
J(\mub_f)&=&[\gb-\Phib\Wb \mub_f]^t<\Bb>[\gb-\Phib\Wb \fb]+(1/\theta_f)\|\Qb\fb\|^2 \\ 
J(\mub_w)&=&[\gb-\Phib\Fb \mub_w]^t<\Bb>[\gb-\Phib\Fb \wb]+\|\Ab\wb\|^2 
\earr
\eeq
Les d\'etails de ces calculs seront omis ici.  

\section{Restauration avec Mod\'elisation Gauss-Markov-Potts}
Le cas d'une mod\'elisation gaussienne reste assez restrictif pour la mod\'elisation des images. 
Des mod\'elisation par des champs de Markov composites (intensit\'es-contours ou intensit\'es-r\'egions) 
sont mieux adapt\'ees. Dans ce travail, nous examinons ce dernier. L'id\'ee de base est 
de classer les pixels de l'images $\fb=\{f(\rb), \rb\in\Rc\}$ en $K$ classes \'etiquett\'ees par une variable discr\`ete $z(\rb)\in\{1,\cdots,K\}$. L'image $z(\rb)=\{f(\rb), \rb\in\Rc\}$ repr\'esente ainsi la segmentation de l'image $f(\rb)$. Chaque paquets des pixels $\fb_k=\{f(\rb), \rb\in\Rc_k\}$ repr\'esente un ensemble fini des r\'egions compacts et disjointes~: 
$\cup_l \Rc_{kl}=\Rc_k$ et $\cup_k \Rc_{k}=\Rc$. 
On suppose aussi que $\fb_k$ et $\fb_l$ $\forall k\not=l$ sont ind\'ependants.  

A chaque r\'egion est associ\'ee un contour. Si on repr\'esente les contours de l'images par 
une variable binaire $q(\rb)$, on a $q(\rb)=0$ \`a l'int\'erieure d'une r\'egion et $q(\rb)=1$ aux fronti\`eres de ces r\'egions. On note aussi que $q(\rb)$ \`a partir de $z(\rb)$ s'obtient d'une mani\`ere d\'eterministe (voir Fig~1). 

Avec cette introduction, nous pouvons d\'efinir
\beq
p(f(\rb)|z(\rb)=k,m_k,v_k)=\Nc(m_k,v_k) 
\eeq
ce qui sugg\`ere un mod\`ele de m\'elange de gaussienne pour les pixels de l'image
\beq
p(f(\rb))=\sum_k a_k \Nc(m_k,v_k) \mbox{~avec~} a_k=P(z(\rb)=k)
\eeq
Une premi\`ere mod\'elisation simple est donc, suppos\'ees que les pixels de l'images sont 
\aprio ind\'ependants, ce qui sugg\`ere
\beq
p(\zb)=\prod_{\rb} p(z(\rb)) 
\eeq
Nous apellons ce mod\`ele, M\'elange de Gaussiennes Ind\'ependantes (MGI). 

\bfig
\btabu{ccc}
\includegraphics[width=25mm,height=25mm]{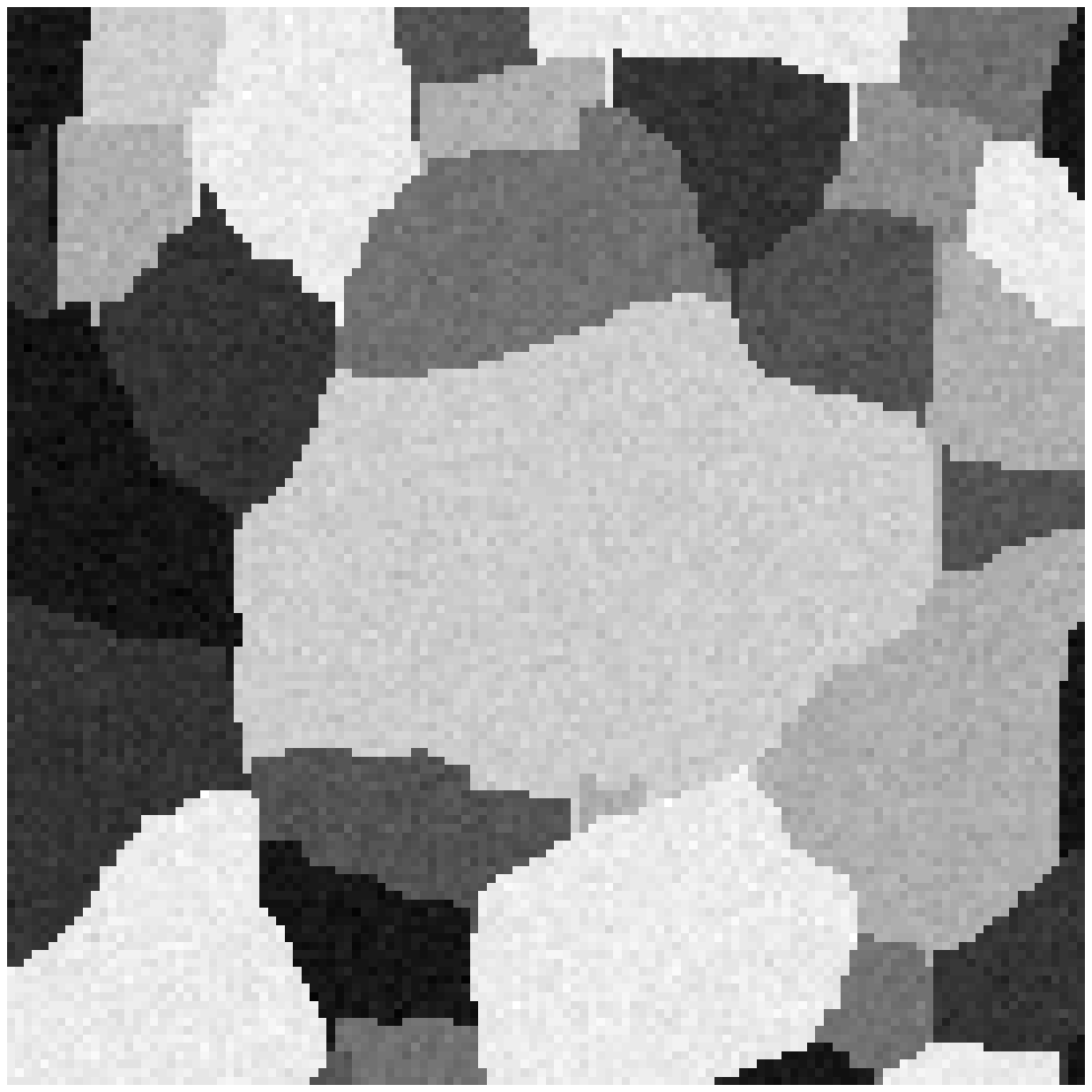}& 
\includegraphics[width=25mm,height=25mm]{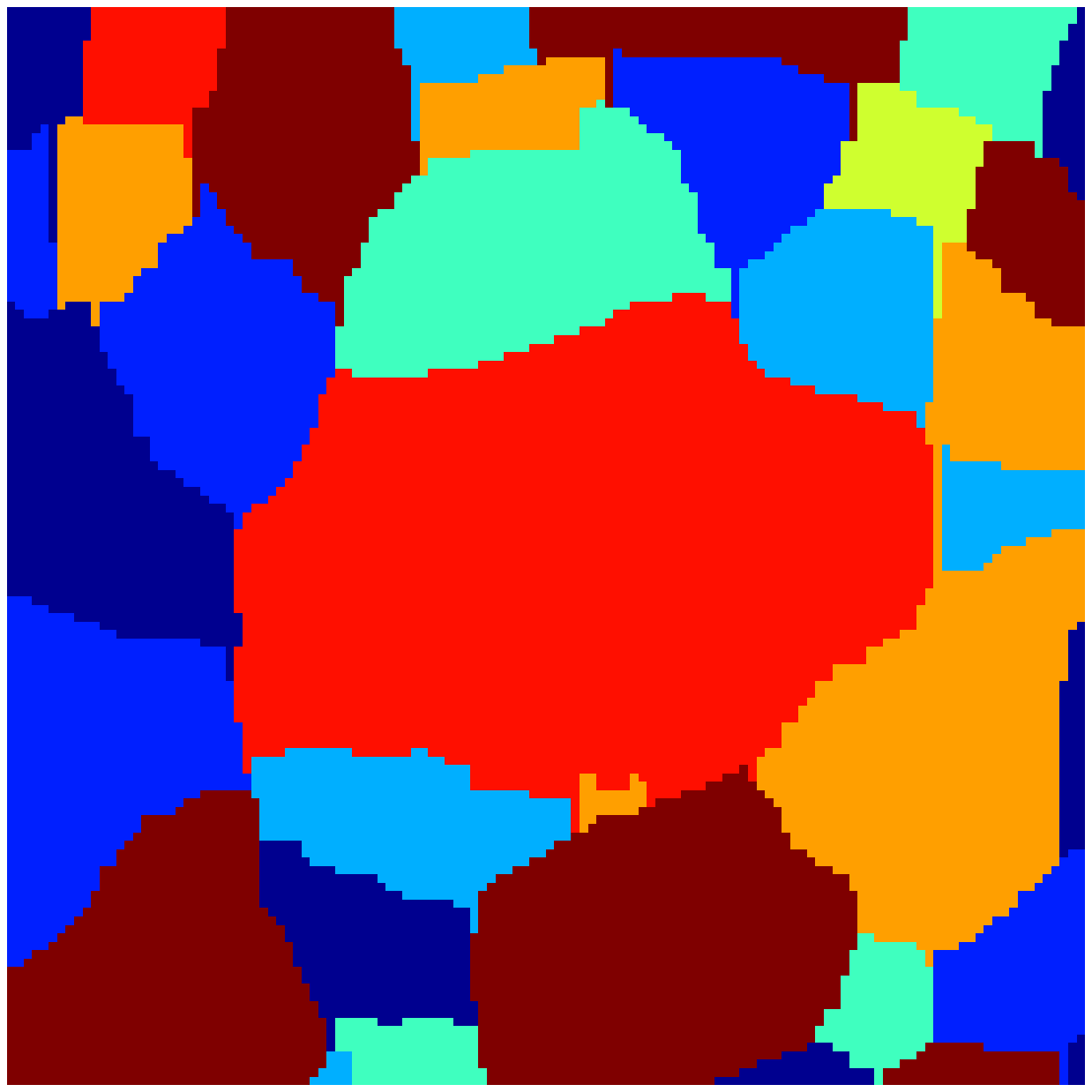}& 
\includegraphics[width=25mm,height=25mm]{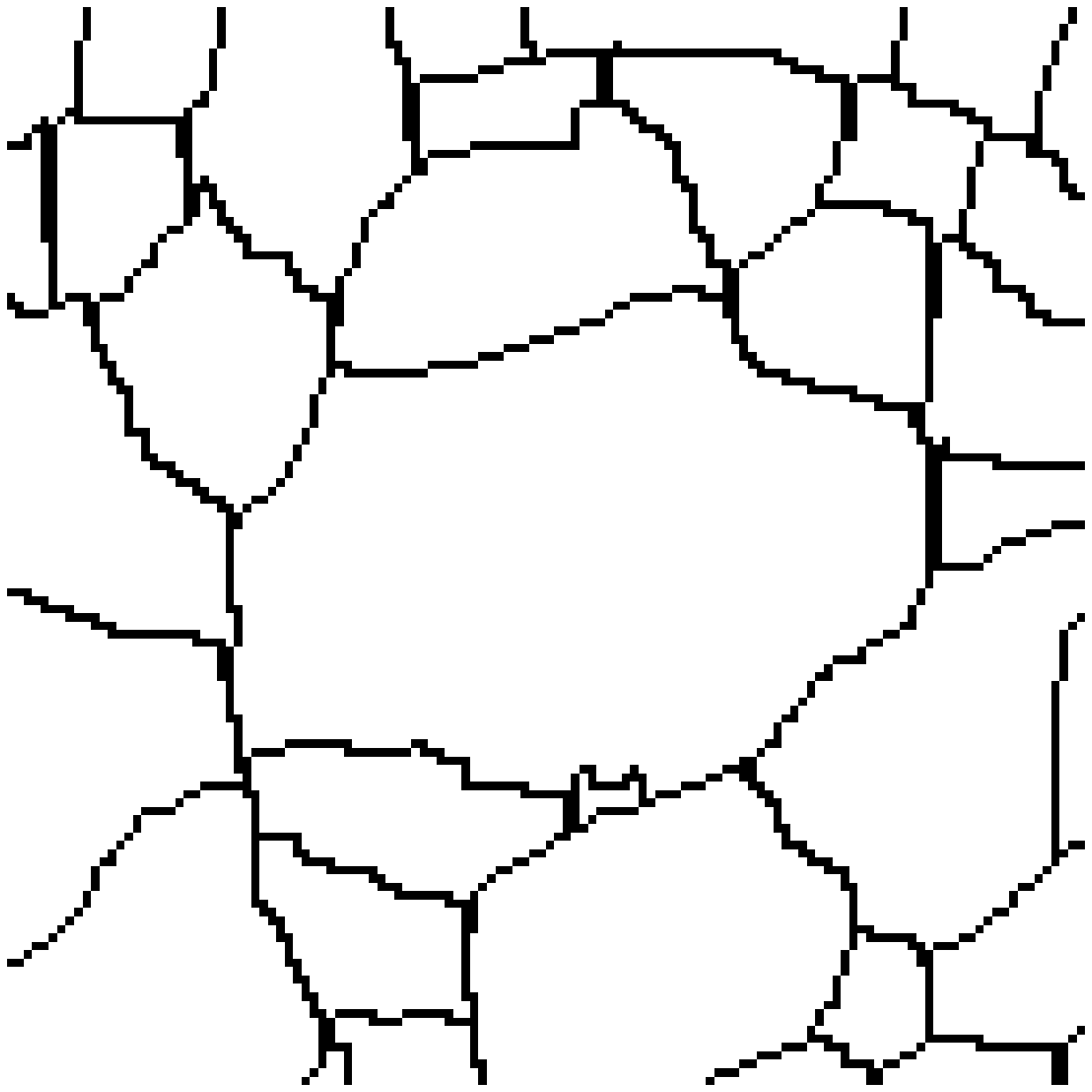}
\\ 
$f(\rb)$ & $z(\rb)$ & $q(\rb)$
\etabu
\caption[Mod\`ele de m\'elange et champs de Markov cach\'e]{Mod\`ele de m\'elange et champs de Markov cach\'e: image des intensit\'es ou niveau de gris $f(\rb)$, image $z(\rb)$ de segmentation ou classification, image binaire $q(\rb)$ des contours.} 
\label{Fig1}
\efig

Maintenant, pour prendre en compte la coh\'erence spatiale de ces pixels, nous devons introduire, d'une mani\`ere ou autre, une d\'ependance spatiale entre ces pixels. La mod\'elisation markovienne est justement l'outil appropri\'e. 

Cette d\'ependance spatiale peut \^etre fait de trois mani\`eres. Soit utiliser un mod\`ele markovien pour $z(\rb)$ et un mod\`ele ind\'ependant pour $f(\rb)|z(\rb)$, soit un mod\`ele markovien pour $f(\rb)|z(\rb)$ et un mod\`ele ind\'ependant pour $z(\rb)$, soit un mod\`ele markovien pour $f(\rb)|z(\rb)$ et un mod\`ele markovien aussi $z(\rb)$. 
Nous avons examin\'e ces cas avec des mod\`eles de Gauss-Markov pour $f(\rb)|z(\rb)$ et le mod\`ele de Potts pour $z(\rb)$. Ce dernier peut s'\'ecrire sous deux formes~:
\beq
p(z(\rb)|z(\rb'),\rb'\in\Vc(\rb))\propto 
\expf{\gamma \sum_{\rb'\in\Vc(\rb)}\delta(z(\rb)-z(\rb'))}
\eeq
\beq 
p(\zb)\propto 
\expf{\gamma \sum_{\rb\in\Rc}\sum_{\rb'\in\Vc(\rb)}\delta(z(\rb)-z(\rb'))}
\eeq
Ces diff\'erents cas peuvent alors se r\'esumer par~:

\noindent{\bf Mod\`ele Gauss-Potts~:}
\beq
\barr{l}
p(f(\rb)|z(\rb)=k)=\Nc(m_k,v_k), \forall \rb\in\Rc \\ 
p(\fb|\zb)=\prod_{\rb\in\Rc} \Nc(m(\rb),v(\rb))  
\earr
\eeq
avec $m(\rb)=m_k, \forall \rb\in\Rc_k$ et $v(\rb)=v_k, \forall \rb\in\Rc_k$,  
et $p(\zb)$ Potts. 

\noindent{\bf Mod\`ele de m\'elange ind\'ependante de Gauss-Markov~:}
\beq
\barr{l}
p(f(\rb)|f(\rb'),\rb'\in\Vc(\rb),q(\rb,\rb'))=\Nc(m(\rb),v(\rb)), \forall \rb\in\Rc \\ 
p(\fb|\zb)\propto \prod_{k} \Nc(m_k\oneb_k,\Sigmab_k)\\ 
p(\zb)=\prod_{\rb} p(z(\rb)=k)=\prod_k \alpha_k^{\sum_{\rb\in\Rc} \delta(z(\rb)-m_k)}
\earr
\eeq
avec $\oneb_k=1, \forall \rb\in\Rc_k$ et $\Sigmab_k$ une matrice de covariance. 

\noindent{\bf Mod\`ele de Gauss-Markov-Potts~:}
\beq
\barr{l}
p(f(\rb)|f(\rb'),\rb'\in\Vc(\rb),q(\rb,\rb'))=\Nc(m(\rb),v(\rb)), \forall \rb\in\Rc \\ 
p(z(\rb)|z(\rb'),\rb'\in\Vc(\rb))\propto
\expf{\gamma \sum_{\rb'\in\Vc(\rb)}\delta(z(\rb)-z(\rb'))}\\ 
\earr
\eeq

Quelque soit le mod\`ele choisi parmi ces diff\'erents mod\`eles, l'objectif est 
d'estimer $\fb$, $\zb$ et $\thetab$. Si on \'ecrit la loi \apost jointe~:
\beq
p(\fb,\zb,\thetab|\gb)=\frac{p(\gb|\fb,\thetab) \; p(\fb|\zb,\thetab) \; p(\zb)}{p(\gb|\thetab)}
\eeq
et on cherche \`a l'approximer par une loi s\'eparable 
$q(\fb,\zb,\thetab|\gb)=q_1(\fb)\;q_2(\zb)\; q_3(\thetab)$. 

Cependant, ici, nous choisissons d'approximer seulement 
$p(\fb,\zb,\thetab)$ par $q_1(\fb|\zb)\;q_2(\zb)\; q_3(\thetab)$. 
Les d\'etails de ces calculs et une comparaison de ces diff\'erents algorithmes sont en cours d'exp\'erimentation et de r\'edaction et seront publi\'es dans un d\'elai tr\`es proche.  

\section{Conclusion}
L'approche variationelle de l'approximation d'une loi par des lois s\'eparables  est appliqu\'ee au 
cas de l'estimation non supervis\'ee des inconnues et des hyper-param\`etres dans des probl\`emes 
inverses de restauration d'image (d\'econvolution simple ou aveugle) avec des mod\'elisations \aprio gaussiennes, 
m\'elange de gaussiennes ou m\'elange de gaussiennes avec champ de labels markovien (champs de Markov cach\'e).  

{\small 
\bibliographystyle{ieeetr}
\def\bibdira{/home/djafari/Tex/Inputs/bib/commun/}
\def\bibdir{/home/seismic/TeX/biblio/}
\def\UP#1{\uppercase{#1}}
\bibliography{\bibdira revuedef,\bibdira bibdef,\bibdir baseAJ,\bibdir baseKZ,\bibdira gpibase,varBayes}
}

\end{document}